\documentclass[]{raa}
\usepackage[noblocks]{authblk}
\usepackage{cases}            
\usepackage{graphicx,times}
\usepackage{url}
\usepackage{pdflscape}
\usepackage{longtable}
\usepackage{multirow}

\begin{document}

\title{The Intrinsic $\gamma$-ray Emissions of $Fermi$ Blazars}

\volnopage{ {\bf 201X} Vol.\ {\bf XX} No. {\bf XX}, 000--000}
\setcounter{page}{1}

\author{C. Lin
  \inst{1,2} \footnote{\it linchao@e.gzhu.edu.cn}
\and J. H. Fan
  \inst{1,2} \footnote{Corresponding author: {\it fjh@gzhu.edu.cn}}
\and H. B. Xiao
  \inst{1,2}
  }

\institute{Center for Astrophysics, Guangzhou University,  Guangzhou 510006, China;
    \and
    Astronomy Science and Technology Research Laboratory of Department of Education of Guangdong Province, Guangzhou 510006, China \vs \no \\
    {\small Received [year] [month] [day]; accepted [year] [month] [day]}
}

\abstract{
Beaming effect is important for the observational properties of blazars. In this work, we collect 91 $Fermi$ blazars with available radio Doppler factors. $\gamma$-ray Doppler factors are estimated and compared with radio Doppler factors for some sources. The intrinsic (de-beamed) $\gamma$-ray flux density ($f^{\rm in}_{\gamma}$), intrinsic $\gamma$-ray luminosity ($L^{\rm in}_{\gamma}$), and intrinsic synchrotron peak frequency ($\nu_{\rm p}^{\rm in}$) are calculated. Then we study the correlations between $f^{\rm in}_{\gamma}$ and redshift and find that they follow the theoretical relation: $\log f = -2.0 \log z + {\rm const}$. When the subclasses are considered, we find that stationary jets are perhaps dominant in low synchrotron peaked blazars. 63 $Fermi$ blazars with both available short variability time scales ($\Delta T$) and Doppler factors are also collected. We find that the intrinsic relationship between $L ^{\rm in}_{\gamma}$ and $\Delta T^{\rm in}$ obeys the Elliot \& Shapiro and the Abramowicz \& Nobili relations. Strong positive correlation between $f_{\gamma}^{\rm in}$ and $\nu_{\rm p}^{\rm in}$ is found, suggesting that synchrotron emissions are highly correlated with $\gamma$-ray emissions.
}

\authorrunning{C. Lin, J. H. Fan \& Xiao H. B.}            
\titlerunning{The Intrinsic $\gamma$-ray Emissions of $Fermi$ Blazars}  
   \maketitle


%
%
\section{Introduction}  
\label{sect:intro}
Two major subclasses of active galactic nuclei (AGNs) are radio loud AGNs and radio quiet AGNs. The radio-loud AGNs are blazars that have high and variable polarization, rapid and high amplitude variability, superluminal motions, and strong $\gamma$-ray emissions, etc. Blazars have two subclasses, namely flat spectrum radio quasars (FSRQs) and BL Lacertae objects (BL Lacs). The difference between the two subclasses is mainly that  BL Lacertae objects show very weak (or even no) emission line features while FSRQs display strong emission lines. However, the continuum emission properties of BL Lacs and FSRQs are quite similar (Zhang \& Fan 2003;
Fan et al. 2009, 2014a;
Xiao et al. 2015).
BL Lac objects were separately discovered through radio or X-ray surveys, and were divided into radio selected BL Lacertae objects (RBLs) and X-ray selected BL Lacertae objects (XBLs). From spectral energy distributions (SEDs), blazars can be divided into
low synchrotron peaked (LSP, $\nu_{\rm{peak}}^{s} <10^{14}$ Hz),
 intermediate synchrotron peaked  (ISP, $10^{14}$ Hz $<\nu_{\rm{peak}}^{s} < 10^{15}$ Hz), and
 high synchrotron peaked (HSP, $\nu_{\rm{peak}}^{s} > 10^{15}$ Hz) blazars. This classification was first proposed by
  Abdo et al. (2010a) (see also
  Wu et al. 2007;
  Yang et al. 2014;
  Ackermann et al. 2015;
  Fan et al. 2015a;
  Lin \& Fan 2016).
In our recent work (Fan et al. 2016), a sample of 1392 \emph{Fermi} blazars is collected, and SEDs of them are obtained. Then we followed the acronym of Abdo et al. (2010a), proposed a classification by using the Bayesian classification method, as follows: $\nu_{\rm peak}^{\rm s}<10^{14}$ Hz for LSP, $10^{14}$ Hz $<\nu_{\rm peak}^{\rm s}<10^{15.3}$ Hz for ISP, and $\nu_{\rm peak}^{s}>10^{15.3}$ Hz for HSP. We also pointed out that there is no ultra-HSP blazars (Fan et al. 2016).

 Blazars have strong $\gamma$-ray emissions, some of them even were detected in TeV energy region
(Weekes T. C. 1997;
Catanese \& Weekes 1999;
Holder, 2012;
Xiong et al. 2013;
Lin \& Fan 2016).
But the origin of high energy emissions are still unclear. The \emph{Fermi} Large Area Telescope (LAT) was launched in 2008, and detected many blazars at $\gamma$-ray energy region
(Abdo et al. 2010b;
   Nolan et al. 2012;
   Acero et al. 2015;
   Ackermann et al. 2015).
   Compared with the predecessor the \emph{Energetic Gamma Ray Experiment Telescope} (\emph{EGRET}), the \emph{Fermi}/LAT satellite has unprecedented sensitivity in the $\gamma$-ray band (Abdo et al. 2010a). The 3rd \emph{Fermi} Large Area Telescope source catalog (3FGL) contains 3033 sources (see Acero et al. 2015), which gives us a large sample to analyze the mechanism of $\gamma$-ray emissions and other observed properties for blazars.

Beaming effect is included in the explanations of the whole electromagnetic emissions including $\gamma$-ray emissions for blazars, and many authors found that their $\gamma$-ray emissions have strong beaming effect (eg., Fan et al. 1999a, 2013a, 2014b, c, 2015b;
Fan, 2005;
Ruan et al. 2014;
Chen et al. 2015, 2016;
Cheng et al. 1999).
Correlations are found between $\gamma$-ray emissions and other bands, and gamma-ray loud blazars are found to have larger Doppler
factors than non-gamma-ray loud ones
(Dondi \& Ghisellini 1995;
Valtaoja \& Ter\"asranta 1995;
Xie et al. 1997;
Fan et al. 1998;
Cheng et al. 2000;
Jorstad et al. 2001a, b;
L\"ahteenm\"aki \& Valtaoja 2003;
Kellermann et al. 2004;
Lister et al. 2009;
Savolainen et al. 2010;
Zhang et al. 2012;
Li et al. 2013;
Xiong et al. 2013;
Wu et al. 2014;
Xiao et al. 2015;
Chen et al. 2016;
Pei et al. 2016).

In a beaming model, the relativistic jet emissions are boosted, $f^{ \rm {ob}} = \delta^q f^{ \rm{in}}$, here $f^{\rm{in}}$ is the intrinsic (de-beamed) emissions in the source rest frame, $\delta$ is a Doppler boosting factor, $q$ depends on the shape of the jet: $q = 2 + \alpha$ for the stationary jet, $q = 3 + \alpha$ for the jet with distinct ``blobs'', and $\alpha$ is an energy spectral index ($f_{\nu} \propto \nu^{-\alpha}$). The Doppler boosting factor, which is defined as $\delta = [\Gamma (1- \beta \cos \theta )]^{-1}$, is important to investigate the intrinsic properties of blazars. But it depends on two unobservable factors: the bulk Lorentz factor, $\Gamma=(1-\beta^2)^{-1/2}$, and the viewing angle, $\theta$, here $\beta$ is the jet speed in units of the speed of light (see Fan et al. 2009; Savolainen et al. 2010).

Some methods are proposed to estimate the Doppler factors. Ghisellini et al. (1993) gave a method of estimating the Doppler factors, which was based on the synchrotron self-Compton model. This method assumes the X-ray flux originates from the self-Compton components, so a predicted X-ray flux can be calculated by using \emph{Very Long Baseline Interferometry} (\emph{VLBI}) data. By comparing this to the observed X-ray flux, the Doppler boosting factors can be calculated. After that, L\"ahteenm\"aki \& Valtaoja (1999, hereafter LV99) proposed a more accurate and reliable method: the variability brightness temperature of the source ($T_{\rm b}^{\rm ob}$) obtained from the variability of VLBI data is used to compare with intrinsic brightness temperature of the source ($T_{\rm b}^{\rm in}$). $T_{\rm b}^{\rm in}$ is assumed to be the equipartition brightness temperature ($T_{\rm b}^{\rm{eq}}$), namely $T_{\rm b}^{\rm in} = T_{\rm b}^{\rm{eq}} = 5 \times 10^{10}$K. So the Doppler factor can be estimated by using $\delta = (T_{\rm b}^{\rm ob}/T_{\rm b}^{\rm in})^{1/3}$. When the variability timescales are obtained from total flux density (TFD) observation, the variability brightness temperature can be calculated by using exponential flares and the variability timescales (LV99, see also Fan et al. 2009; Hovatta et al. 2009).

Because of the short term variability, a highly compact engine exists at the center of blazars. The balance between gravitation and radiation pressure determines an upper limit of luminosity, namely Eddington luminosity, for any active galactic nuclei (Bassani et al. 1983). If the short variability time scale is assumed to be equal or greater than the time that light travels across the Schwarzschild radius of a black hole, then the observed luminosity and short variability time scale should obey to the so called Elliot \& Shapiro Relation (E-S Relation): $ \rm log L \leq 43.1 + log \,\, \Delta T$ (Elliot \& Shapiro 1974), where $L$ is luminosity in units of erg/s, and $\Delta T$ is variability time scale in units of second (s).
Generally, short variability time scale is assumed to be the time scale which is shorter than one week (Fan 2005).
When the anisotropy of emissions is considered, above limit was replaced by the Abramowicz \& Nobili Relation (A-N Relation): $ \rm log L \leq 44.3 + log \,\, \Delta T$ (Abramowicz \& Nobili 1982).

The intrinsic properties are required to analyze the origin of $\gamma$-ray emissions for blazars. In our recent work of Xiao et al. (2015), we considered the beaming effect of \emph{Fermi} blazars in Nolan et al. (2012) (2FGL), then analysed the correlation between $\gamma$-ray flux density and redshift for 73 blazars, and the relation between $\gamma$-ray short variability time scale and $\gamma$-ray luminosity by comparing with the E-S Relation and the A-N Relation for 28 blazars. In this work, we use a larger sample to revisit the relation between $\gamma$-ray flux density and redshift, and the relation between $\gamma$-ray luminosity and short variability time scale. The subclasses of blazars, and the short variability time scale from X-ray and optical bands are also considered. Then we have 91 \emph{Fermi} blazars with available radio Doppler factors and 63 \emph{Fermi} blazars with both available short variability time scales and Doppler factors. $\gamma$-ray Doppler factors are estimated for the \emph{Fermi} blazars with available short variability time scales at optical, X-ray or $\gamma$-ray band. In addition, the correlation between $\gamma$-ray emissions and synchrotron peaked frequency is also studied in this work. This work is arranged as follows: we will describe our sample and corresponding results in section 2, and give some discussions and conclusions in section 3.

\section{Sample and Results}

\subsection{Sample}
 In this work, we collect blazars with available radio Doppler factors from the literatures: LV99, Fan et al. (2009), Hovatta et al. (2009) and Savolainen et al. (2010). These references used the same method introduced by LV99. If the Doppler factors of some sources are available in more than one reference, we choose the value from the latest paper. Based on the third catalog of active galactic nuclei detected by the Fermi Large Area Telescope (3LAC)\footnote{http://www.asdc.asi.it/fermi3lac/} (Ackermann et al. 2015), we collect a sample of 91 \emph{Fermi} blazars with available radio Doppler factors, which are listed in Table 1.
In Table 1, $\gamma$-ray data are from 3LAC, only that for PKS 2145+06 is from 2LAC\footnote{http://www.asdc.asi.it/fermi2lac/} (Ackermann et al. 2011).

We get the \emph{Fermi} integral photon flux at 1--100 GeV, as we did in our previous works (Fan et al. 2013a, 2014b), let
$${\frac{dN}{dE}} = N_{0} E^{-\alpha^{{\rm ph}}},$$
where ${\rm\alpha^{ph}}$ is a photon spectral index, here $\alpha^{{\rm ph}}= \alpha -1$. Then the flux density at energy $E_0$ in units of GeV/cm$^2$/s/GeV can be expressed as
\begin{equation}
f_E=\frac{N_{(E_{\rm L}\sim E_{\rm U})}(1-\alpha^{{\rm ph}})}{E_{\rm U}^{1-\alpha^{{\rm ph}}}-{E_{\rm L}^{1-\alpha^{{\rm ph}}}}}\cdot E_0^{1-\alpha^{{\rm ph}}},
\end{equation}
where $ N_{(E_{\rm L}\sim E_{\rm U})}$ is integral photons in units of ${\rm ph/cm^2/s}$ in the energy range of $ E_{\rm L}\sim E_{\rm U}$.
Because the integral flux in GeV/cm$^2$/s can be obtained by $F=\int  ^{E_{\rm U}}_{E_{\rm L}} EdN=\int  ^{E_{\rm U}}_{E_{\rm L}} f_E dE$, therefore

$$
F= \frac{N_{(E_{\rm L}\sim E_{\rm U})}E_{\rm U} \times E_{\rm L}}{E_{\rm U}-E_{\rm L}} \ln (\frac{E_{\rm U}}{E_{\rm L}}), \,\,{\rm for\,\, \alpha^{ph}=2}
$$

\begin{equation}
F= N_{(E_{\rm L}\sim E_{\rm U})}{\frac{1-\alpha^{\rm ph}}{2-\alpha^{\rm ph}}}{\frac{(E_{\rm U}^{2-\alpha^{\rm ph}}-E_{\rm L}^{2-\alpha^{\rm ph}})}{(E_{\rm U}^{1-\alpha^{\rm ph}}-E_{\rm L}^{1-\alpha^{\rm ph}})}}, \,\,{\rm otherwise.}
\end{equation}
For the \emph{Fermi} sources in this work, $E_{\rm L}$ and $E_{\rm U}$ correspond to 1 GeV and 100 GeV respectively.

The synchrotron peak frequency (${\rm log\nu^{s}_{p}}$) of PKS 2145+06 is not available in Fan et al. (2016) or 2LAC. We use the empirical relationship  introduced in Fan et al. (2016) to estimate it, as follows
\begin{equation}
\log \nu_{\rm p}^{\rm s}=
\left\lbrace
  \begin{array}{ll}
     16 + 4.238X  & \,\,\,\,\,\,\,X<0 \\
     16 + 4.005Y  & \,\,\,\,\,\,\,X>0,
  \end{array}\right.
\end{equation}
where $X = 1 - 1.262 \alpha_{\rm RO} - 0.623 \alpha_{\rm OX} $, and $Y = 1.0 + 0.034 \alpha_{\rm RO} - 0.978 \alpha_{\rm OX}$ (Fan et al. 2016), $\alpha_{\rm RO}$ and $\alpha_{\rm OX}$ are the effective spectral indexes. For PKS 2145+06, we can get $\alpha_{\rm RO}=0.666$ and $\alpha_{\rm OX}=1.283$ from 2LAC, so we obtain $\log\nu_{\rm p}^s=13.29$ Hz.

\subsection{$\gamma$-ray Flux Density and Redshift}
Our sample in Table 1 contains 32 BL Lacs and 59 FSRQs, or 40 ISP and 51 LSP based on the SED classification in Fan et al. (2016). In this work, we transform the $\gamma$-ray integral photon flux at 1--100 GeV into the flux density in units of mJy at $E_0=2$GeV by using the equation (1), and adopt a linear regression to the correlation between flux density and redshift for the whole sample, BL Lacs, FSRQs, ISP, and LSP respectively. All the fluxes are K-corrected by using $f=f^{\rm obs}(1+z)^{\alpha-1}$.

\begin{figure}[h]
  \centering
  \includegraphics[width=5.00in]{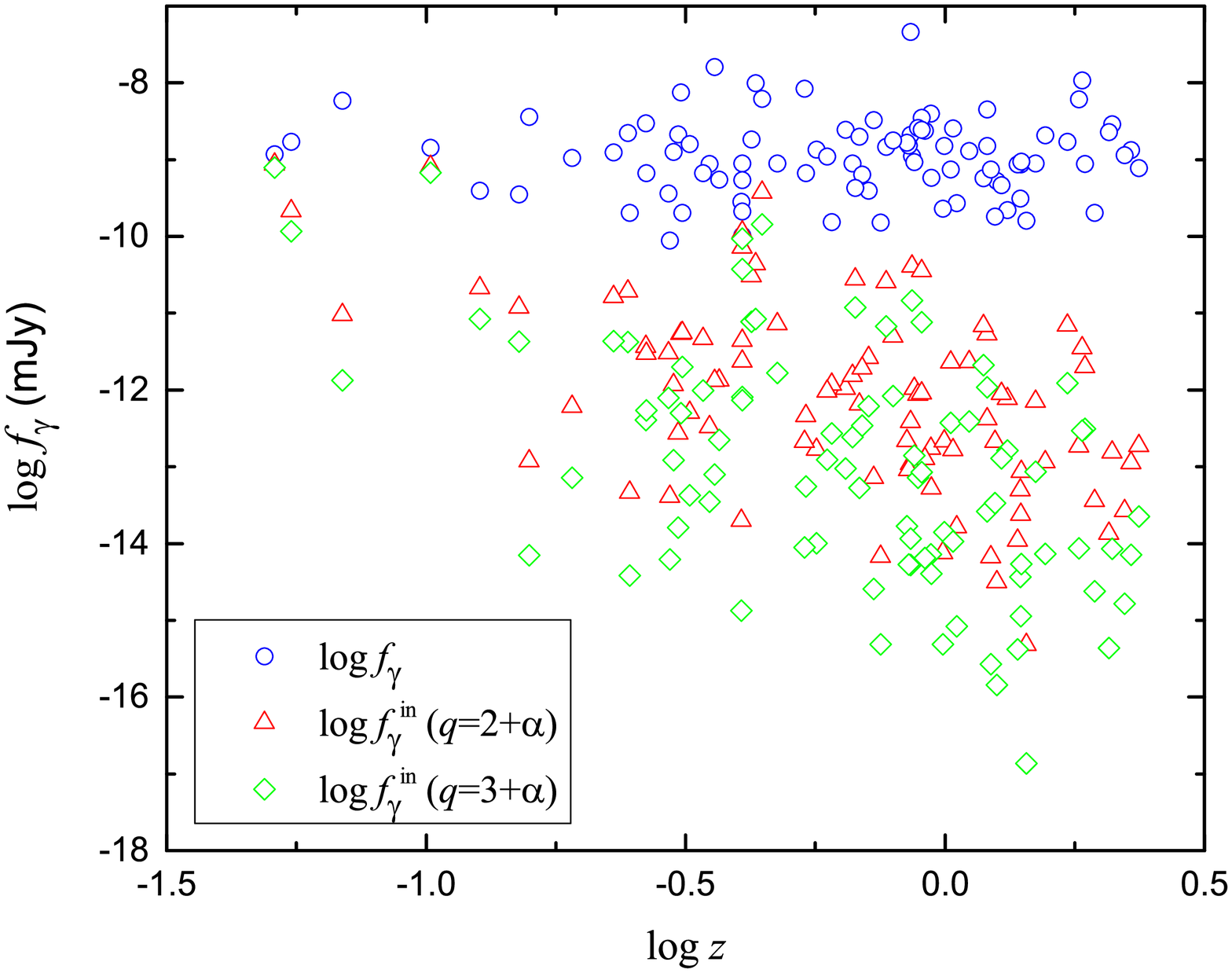}\\
  \caption{Polt of the $\gamma$-ray flux density against the redshift for the whole 91 blazars. The circle symbols stand for the observed values, the triangle symbols stand for the intrinsic values estimated in the case of $q=2+\alpha$, the rhombus symbols stand for the intrinsic values estimated in the case of $q=3+\alpha$.}
  \label{Lin-2015-intrinsic-logf-logz}
\end{figure}

{\it \underline{Whole sample:}} For the whole sample of 91 blazars, we have
$
   \log f_{\gamma} = - (0.01 \pm 0.15) \log z - (8.96 \pm 0.06)
$
with a correlation coefficient $r = -0.01 $ and a chance probability $p = 93.08\%$.
As we introduce in the Section 1, we can calculate the intrinsic flux density, then we have
$
   \log f_{\gamma}^{\rm in} = - (1.82 \pm 0.30) \log z - (12.44 \pm 0.13)
$
with $r = -0.54 $ and $p = 2.98 \times 10^{-8}$ for the case of $q=2+\alpha$, and
$
   \log f_{\gamma}^{\rm in} = - (2.32 \pm 0.37) \log z - (13.46 \pm 0.16)
$
with $r = -0.55 $ and $p = 1.66 \times 10^{-8}$ for $q=3+\alpha$. The corresponding figure is shown in Fig. \ref{Lin-2015-intrinsic-logf-logz}.

\begin{figure}[h]
  \centering
  \includegraphics[width=4.00in]{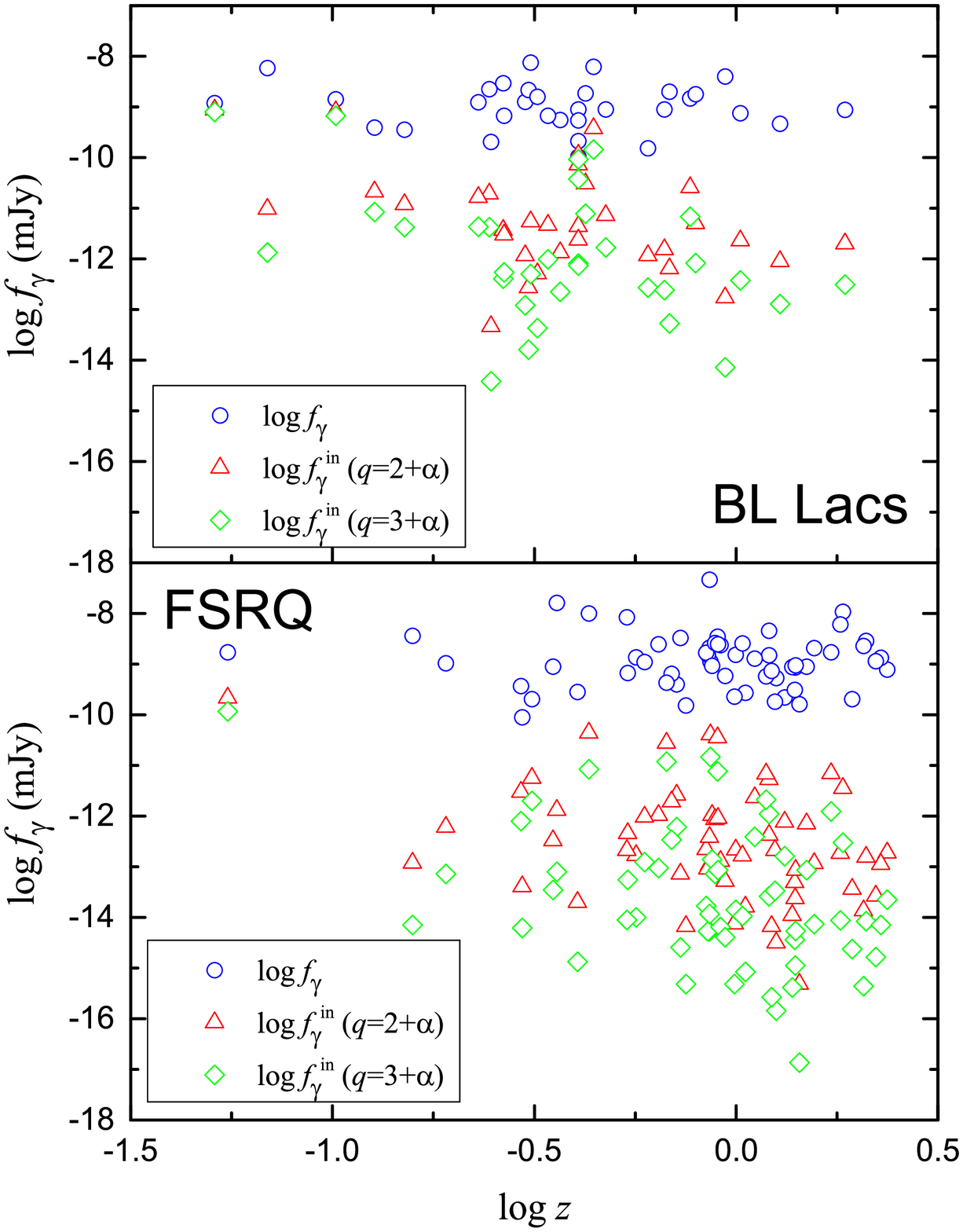}\\
  \caption{Polt of the $\gamma$-ray flux density against the redshift for 32 BL Lacs (upper panel) and 59 FSRQs (lower panel). The circle symbols stand for the observed values, the triangle symbols stand for the intrinsic values estimated in the case of $q=2+\alpha$, the rhombus symbols stand for the intrinsic values estimated in the case of $q=3+\alpha$.}
  \label{Lin-2015-intrinsic-logF-logz-BQ}
  \end{figure}

{\it \underline{BL Lacs:}} For the 32 BL Lacs, we have
$
   \log f_{\gamma} = - (0.13 \pm 0.24) \log z - (9.05 \pm 0.13)
$
with $r = -0.10 $ and $p = 59.96\%$;
$
   \log f_{\gamma}^{\rm in} = - (1.30 \pm 0.47) \log z - (11.82 \pm 0.26)
$
with $r = -0.45 $ and $p = 1.03\%$ for $q=2+\alpha$; and
$
   \log f_{\gamma}^{\rm in} = - (1.65 \pm 0.61) \log z - (12.68 \pm 0.34)
$
with $r = -0.44 $ and $p = 1.13\%$ for $q=3+\alpha$. The corresponding figure is shown in the upper panel of Fig. \ref{Lin-2015-intrinsic-logF-logz-BQ}

{\it \underline{FSRQs:}} For the 59 FSRQs, we have
$
   \log f_{\gamma} = -(0.02 \pm 0.23) \log z - (8.94 \pm 0.07)
$
with $r = -0.01 $ and $p = 94.20\%$;
$
   \log f_{\gamma}^{\rm in} = - (1.34 \pm 0.45) \log z - (12.61 \pm 0.14)
$
with $r = -0.37 $ and $p = 3.90\times10^{-3}$ for $q=2+\alpha$; and
$
   \log f_{\gamma}^{\rm in} = - (1.73 \pm 0.55) \log z - (13.68 \pm 0.17)
$
with $r = -0.39 $ and $p = 2.50\times10^{-3}$ for $q=3+\alpha$. The corresponding figure is shown in the lower panel of Fig. \ref{Lin-2015-intrinsic-logF-logz-BQ}.

\begin{figure}[h]
  \centering
  \includegraphics[width=4.00in]{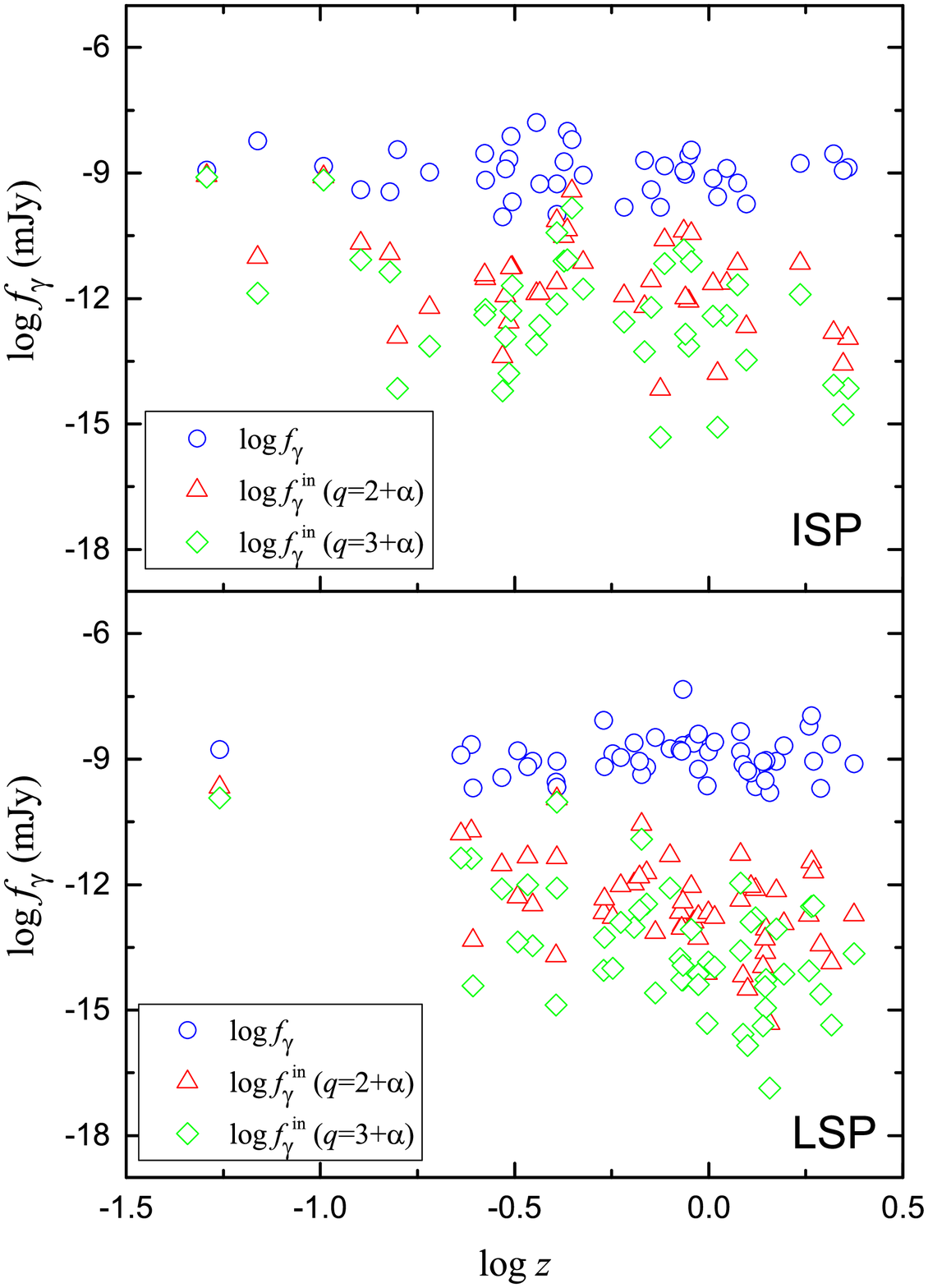}\\
  \caption{Polt of the $\gamma$-ray flux density against the redshift for 40 ISP blazars (upper panel) and 51 LSP blazars (lower panel). The circle symbols stand for the observed values, the triangle symbols stand for the intrinsic values estimated in the case of $q=2+\alpha$, the rhombus symbols stand for the intrinsic values estimated in the case of $q=3+\alpha$.}
  \label{Lin-2015-intrinsic-logF-logz-SED}
\end{figure}

{\it \underline{ISP:}} For the 40 ISP blazars, we have
$
   \log f_{\gamma} = - (0.12 \pm 0.22) \log z - (9.02 \pm 0.11)
$
with $r = -0.09 $ and $p = 58.81\%$;
$
   \log f_{\gamma}^{\rm in} = - (1.33 \pm 0.44) \log z - (12.01 \pm 0.22)
$
with $r = -0.44 $ and $p = 4.12\times10^{-3}$ for $q=2+\alpha$; and
$
   \log f_{\gamma}^{\rm in} = - (1.67 \pm 0.54) \log z - (12.89 \pm 0.27)
$
with $r = -0.45 $ and $p = 3.66\times10^{-3}$ for $q=3+\alpha$. The corresponding figure is shown in the upper panel of Fig. \ref{Lin-2015-intrinsic-logF-logz-SED}.

{\it \underline{LSP:}} For the 51 LSP blazars, we have
$
   \log f_{\gamma} = (0.09 \pm 0.22) \log z - (8.94 \pm 0.07)
$
with $r = 0.06$ and $p = 68.03\%$;
$
   \log f_{\gamma}^{\rm in} = -(1.95 \pm 0.44) \log z - (12.66 \pm 0.14)
$
with $r = -0.54 $ and $p = 5.20 \times 10^{-5}$ for $q=2+\alpha$; and
$
   \log f_{\gamma}^{\rm in} = - (2.51 \pm 0.55) \log z - (13.75 \pm 0.18)
$
   with $r = -0.55 $ and $p = 3.40 \times 10^{-5}$ for $q=3+\alpha$. The corresponding figure is shown in the lower panel of Fig. \ref{Lin-2015-intrinsic-logF-logz-SED}.

\subsection{Short Variability Time Scale and Luminosity}
Observations suggest that $\gamma$-ray loud blazars are variable on time scales of hours although there is no preferred scale for the variation time of any source (Fan et al. 2014a). For example, \emph{Fermi} LAT detected a variability time scale of $\sim$12 hour for PKS $1454-354$ (Abdo et al. 2009), and a doubling time of roughly 4 hours for PKS 1502+105 (Abdo et al. 2010c). In the literatures, available short variability time scales are collected, eg., Bassani et al. (1983), Dondi \& Ghisellini (1995), Fan et al. (1999a), Gupta et al. (2012), Vovk \& Neronov et al. (2013).

For the sources with available short variability time scales, X-ray, and $\gamma$-ray emissions, we can estimate their $\gamma$-ray Doppler factors. Follow our recent work (Fan et al. 2013b, 2014a), a Doppler factor can be estimated using:
$$ \delta_\gamma \ge \left[1.54\times10^{-3}(1+z)^{4+2\alpha}
   \left(\frac{d_\mathrm{L}}{\mathrm{Mpc}}\right)^2
  \left(\frac{\Delta T}{\mathrm{hr}}\right)^{-1}
  \left(\frac{F_\mathrm{KeV}}{\mu\mathrm{Jy}}\right)
  \left(\frac{E_\gamma}{\mathrm{GeV}}\right)^\alpha \right]^{\frac{1}{4+2\alpha}},
  $$
here $\Delta T$ is the variability time scale in units of hrs, $\alpha$ is the X-ray spectral index, $F_\mathrm{KeV}$ is the flux density at 1 KeV in units of $\mu\mathrm{Jy}$, $E_{\gamma}$ is the energy in units of GeV at which the $\gamma$-rays are detected, and $d_\mathrm{L}$ is the luminosity distance in units of Mpc. The average energy $E_{\gamma}$ can be calculated by $E_{\gamma} = \int EdN / \int dN$, and the luminosity distance can be expressed in the form
$$d_{\rm L} =
 (1+z)\frac{c}{H_{0}}\int^{1+z}_{1}\frac{1}{\sqrt{\Omega_{M}x^{3}+1-\Omega_{M}}} \,\,{\rm d}x$$
from the $\Lambda$-CDM model (Capelo \& Natarajan 2007) with $\Omega_{\Lambda}\simeq0.7$, $\Omega_{\rm M}\simeq0.3$ and $\Omega_{\rm K}\simeq0.0$. We adopt $H_0 = 73$ Km/s/Mpc throughout the paper. Then, we estimate the $\gamma$-ray Doppler factors ($\delta_{\gamma}$) of 63 blzazrs, and show them in Table 2.

\begin{figure}[h]
  \centering
  \includegraphics[width=5.00in]{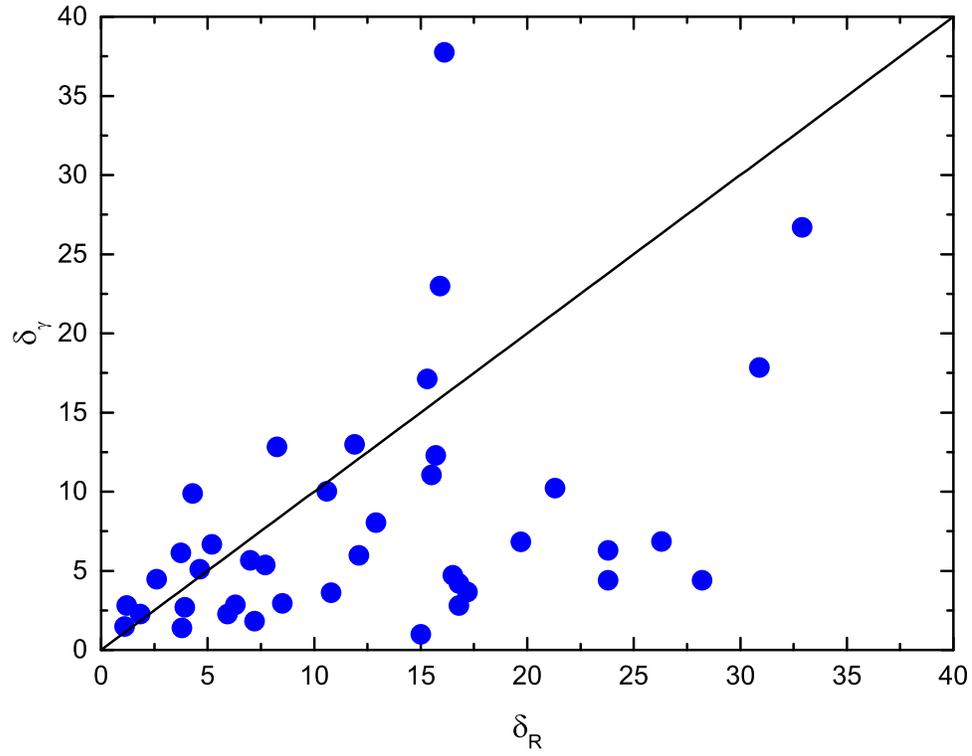}\\
  \caption{Plot of the $\gamma$-ray Doppler factors estimated in this work against the radio Doppler factors from corresponding references.}
  \label{Lin-2015-intrinsic-delta_G-delta_R}
\end{figure}

 To compare $\gamma$-ray Doppler factors estimated in this work with radio Doppler factors, we show the plot of $\gamma$-ray Doppler factors against radio Doppler factors in Fig. \ref{Lin-2015-intrinsic-delta_G-delta_R}. We find that the $\gamma$-ray Doppler factors that we estimated are on average lower than the radio Doppler factors. The reason maybe from the fact that the derived value in this work is a lower value as discussed in Fan et al. (2014a).

The luminosity can be calculated using $ L_{\gamma}=4\pi d_{\rm L}^2 (1+z)^{\alpha-1} F_{\gamma}$, where $F_{\gamma}$ is the integral flux calculated by equation (2), $(1+z)^{\alpha-1}$ stands for a K-correction into the source rest frame (Fan et al. 2013a; Kapanadze 2013). In a beaming model, the observed photon energy are also beamed, $E^{\rm ob}=\delta E^{\rm in}$, here $E^{\rm in}$ is the intrinsic energy.
Because $F = \int f dE$, so we have
 $L^{\rm ob} = \delta^{3+\alpha}L^{\rm in}$ for the case of $q = 2 + \alpha$,
 $L^{\rm ob} = \delta^{4+\alpha}L^{\rm in}$ for the case of $q = 3 + \alpha$,
 and $\Delta T^{\rm in} = \delta \Delta T^{\rm ob}$.
 Here $L^{\rm in}$ and $\Delta T^{\rm in}$ are the intrinsic luminosity and the intrinsic variability time scale respectively.

\begin{figure}[h]
  \centering
  \includegraphics[width=5.00in]{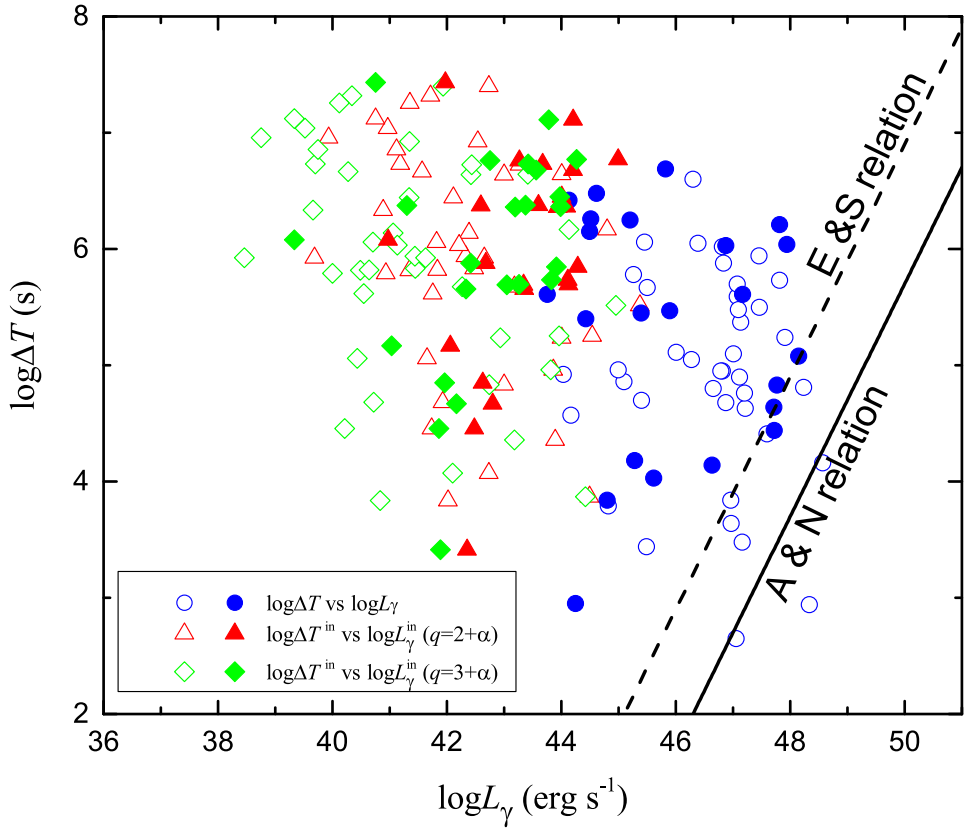}\\
  \caption{Polt of the short variability time scale against the $\gamma$-ray luminosity. The circle symbols stand for the observed values, the triangle symbols stand for the intrinsic values estimated  in the case of  $q=2+\alpha$, the rhombus symbols stand for the intrinsic values estimated in the case of $q=3+\alpha$. The filled signs stand for the sources whose intrinsic values are estimated by the $\gamma$-ray Doppler factors, while the open signs stand for those by the radio Doppler factors.}
  \label{Lin-2015-intrinsic-logT-logL}
\end{figure}

\begin{figure}[h]
  \centering
  \includegraphics[width=4.50in]{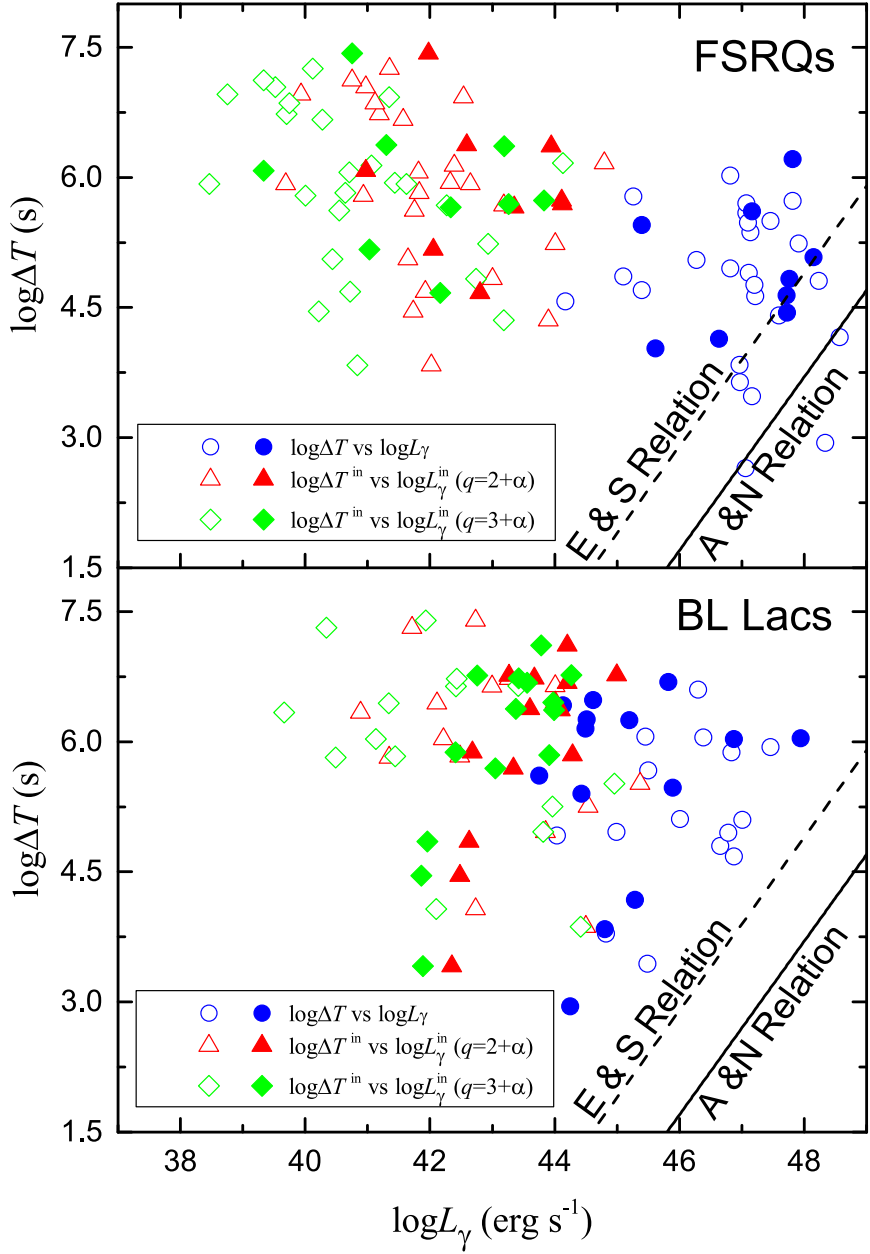}\\
  \caption{Polt of the short variability time scale against the $\gamma$-ray luminosity for 31 FSRQs (upper panel) and 26 BL Lacs (lower panel). The circle symbols stand for the observe values, the triangle symbols stand for the intrinsic values estimated  in the case of  $q=2+\alpha$, the rhombus symbols stand for the intrinsic values estimated in the case of $q=3+\alpha$. The filled signs stand for the sources whose intrinsic values are estimated by the $\gamma$-ray Doppler factors, while the open signs stand for those by the radio Doppler factors.}
  \label{Lin-2015-intrinsic-logT-logL-BQ}
\end{figure}

For calculating $L_{\gamma}^{\rm in}$ and $\Delta T^{\rm in}$, we use radio Doppler factors in Table 1, but if there is no available radio Doppler factor, we used $\gamma$-ray Doppler factors in Table 2 instead. When the Doppler boosting effect is considered, the plot of short variability time scale against the $\gamma$-ray luminosity is shown in Fig. \ref{Lin-2015-intrinsic-logT-logL}, where the observed properties and intrinsic properties are shown respectively. Then we find that 9 blazars violate the E-S or A-N relation in $L_{\gamma}^{\rm ob}$ versus $\Delta T^{\rm ob}$. However, the whole sample follows the E-S and A-N relations in $L_{\gamma}^{\rm in}$ versus $\Delta T^{\rm in}$, see Fig. \ref{Lin-2015-intrinsic-logT-logL}. When the subclasses of blazars are considered, 9 FSRQs violate the E-S relation and 3 FSRQs violate the A-N relation in observed data, while all blazars follow those relations in intrinsic data, see Fig. \ref{Lin-2015-intrinsic-logT-logL-BQ}.

\subsection{$\gamma$-ray Emissions and Synchrotron Peaked Frequency}
For the whole 91 blazars, the linear regression analysis is adopted to the correlations between $\gamma$-ray flux density ($\log f_{\gamma}$) and synchrotron peak frequency ($\log \nu^s_{\rm p}$). The synchrotron peak frequencies are corrected to the rest-frame by $\nu_{\rm p}^{\rm res} = (1 + z) \nu_{\rm p}^{\rm obs}$ before analysis.

\begin{figure}[h]
  \centering
  \includegraphics[width=4.50in]{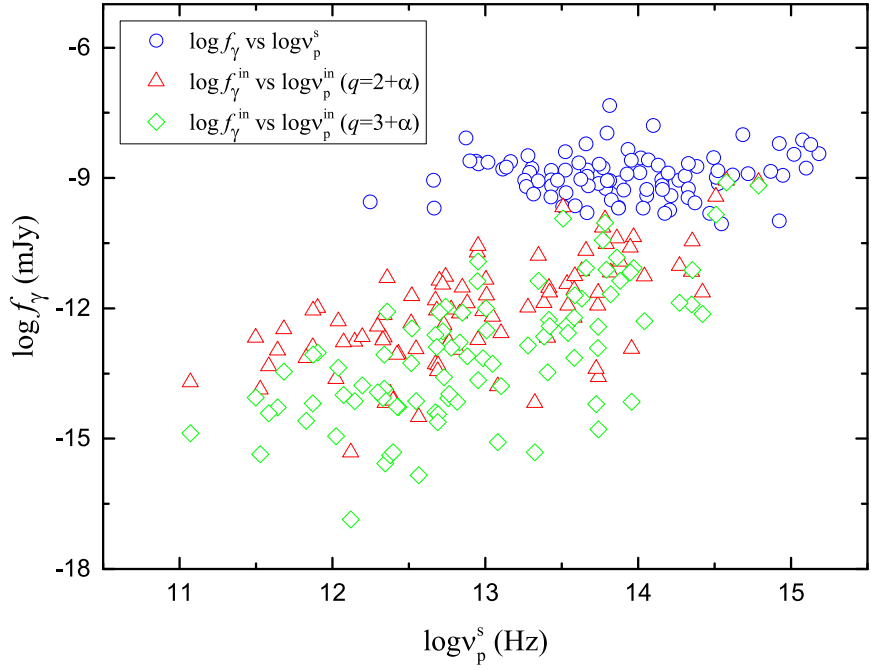}\\
  \caption{Plot of the $\gamma$-ray flux density against the synchrotron peak frequency for the entire 91 blazars. The circle symbols stand for the observed values, the triangle symbols stand for the intrinsic values estimated in the case of $q=2+\alpha$, the rhombus symbols stand for the intrinsic values estimated  in the case of  $q=3+\alpha$.}
  \label{Lin-2015-intrinsic-logF-lognu}
\end{figure}

For $\log f_{\gamma}$--$\log \nu^s_{\rm p}$ correlation, we have
$
   \log f_{\gamma} = (0.06 \pm 0.09) \log \nu^s_{\rm p} - (9.75 \pm 1.21)
$
with $r = 0.07 $ and $p = 51.51\%$,
$
   \log f_{\gamma}^{\rm in} = (1.00 \pm 0.12) \log \nu^{\rm in}_{\rm p} - (25.03 \pm 1.60)
$
with $r=0.65$ and $p=2.54 \times 10^{-12}$ for $q=2+\alpha$, and
$
   \log f_{\gamma}^{\rm in} = (1.30 \pm 0.15) \log \nu^{\rm in}_{\rm p} - (29.84 \pm 1.95)
$
with $r=0.68$ and $p=1.97 \times 10^{-13}$ for $q=3+\alpha$, here $\nu_{\rm p}^{\rm in}=\nu_
{\rm p}^{\rm res}/\delta$ is an intrinsic peak frequency. The corresponding figure is shown in Fig. \ref{Lin-2015-intrinsic-logF-lognu}.

As results shown in Section 2.2 and in Lin \& Fan (2016), $\gamma$-ray flux density is strongly correlated with redshift, therefore the correlations between $\gamma$-ray emissions and peak frequency maybe caused by a redshift effect. In our recent works (Fan et al. 2013a, 2015a; Lin \& Fan 2016), we have removed the redshift effect from luminosity-luminosity correlation by using the partial correlation
introduced by Padovani  (1992).
If variables  $i$ and $j$ are correlated with a third one $k$, then the correlation between $i$ and $j$ can remove the $k$ effect, as follows:
$${r_{ij,k}} = ({r_{ij}} -
{r_{ik}}{r_{jk}})/\sqrt {(1 - r_{ik}^2)(1 - r_{jk}^2)},$$
here $r_{ij}$, $r_{ik}$, $r_{jk}$ are the correlation coefficients between any two variables of $i$, $j$ and $k$ respectively.
When the method is performed to the $\log f_{\gamma}^{\rm in}$--$\log \nu_{\rm p}^{\rm in}$ correlation, then the correlation coefficients excluding the redshift effect for the whole sample are as follows:
 $r_{f \nu, z}$ = 0.56 with $p_{f \nu,z} = 8.09 \times 10^{-9} $ for $q=2+\alpha$, and
 $r_{f \nu, z}$ = 0.59 with $p_{f \nu,z} = 7.52 \times 10^{-10} $ for $q=3+\alpha$.

\section{discussion and conclusion}
In this work, we collect 91 blazars with available radio Doppler factors, and calculate their intrinsic $\gamma$-ray flux density ($\log f _{\gamma}^{\rm in}$) at 2 GeV, and intrinsic synchrotron peak frequency ($\log \nu _{\rm p}^{\rm in}$). Then the correlations between between $\log f_{\gamma}^{\rm in}$ and redshift, and between $\log f_{\gamma}^{\rm in}$ and $\log \nu _{\rm p}^{\rm in}$ are investigated. We also study the intrinsic relation between short variability time scale ($\Delta T^{\rm in}$) and $\gamma$-ray luminosity ($\log L _{\gamma}^{\rm in}$) for 63 blazars by comparing to the E-S and A-N Relations.

In our recent work, we compared the $\gamma$-ray Doppler factors ($\delta_{\gamma}$) with the radio Doppler factors ($\delta_R$), and found they are associated with each other, see Fan et al. (2014a). So that we can investigate the beaming effect in $\gamma$-ray band by using the radio Doppler factors. We also found that the radio Doppler factor well correlates with $\gamma$-ray luminosity for the \emph{Fermi} detected sources, see Fan et al. (2009).
In this work, we find that $\log f_{\gamma}^{\rm in}$ of FSRQs are smaller than that of BL Lacs with a probability for the two groups to come from the same distribution (Kolmogorov-Smirnov test) being $p < 10^{-5}$ for $q=2+\alpha$ and $q=3+\alpha$. The averaged values of $\log f_{\gamma}^{\rm in}$ are $<\log f_{\gamma}^{\rm in}>=-12.52\pm 1.12$ ($q=2+\alpha$), $<\log f_{\gamma}^{\rm in}>=-13.56\pm 1.38$ ($q=3+\alpha$) for FSRQs; and $<\log f_{\gamma}^{\rm in}>=-11.25\pm 1.00$ ($q=2+\alpha$), $<\log f_{\gamma}^{\rm in}>=-11.95\pm 1.29$ ($q=3+\alpha$) for BL Lacs. And a t-test indicates that the averaged difference of $\log f_{\gamma}^{\rm in}$ between BL Lacs and FSRQs is $\Delta (\log f^{\rm in}_{\gamma}) = 1.27 \pm 0.24$ with a significant level of $p < 10^{-6}$ for $q=2+\alpha$, and $\Delta (\log f^{\rm in}_{\gamma}) = 1.61 \pm 0.30$ with $p < 10^{-6}$ for $q=3+\alpha$, while the averaged difference of the observed flux density ($\log f_{\gamma}$) is $\Delta (\log f_{\gamma}) = 0.05 \pm 0.11$ with $p=64.23\%$.

\subsection{$\gamma$-ray Flux Density and Redshift}
We analyse the whole 91 \emph{Fermi} blazars, and find that $\log f_{\gamma}$ is weakly correlated with the redshift ($r=-0.01$, and slope is $-0.01 \pm 0.15$). The result is different from our expectation: $\log f = -2.0 \log z + {\rm const}$, if blazars belong to a group. But when we consider the strong beaming effect of $\gamma$-ray emissions for \emph{Fermi} blazars, we find that $\log f_{\gamma}^{\rm in}$ is strongly correlated with redshift as follows: $\log f_{\gamma}^{\rm in} = - (1.82 \pm 0.30) \log z - (12.44 \pm 0.13)$ with $r = -0.54 $ and $p = 2.98 \times 10^{-8}$ for $q=2+\alpha$, and $\log f_{\gamma}^{\rm in} = - (2.32 \pm 0.37) \log z - (13.46 \pm 0.16)$ with $r = -0.55 $ and $p = 1.66 \times 10^{-8}$ for $q=3+\alpha$. In our recent work (Xiao et al. 2015), we found that $\log f_{\gamma}^{\rm in}$ and redshift have strong correlation ($p < 10^{-4}$) with the slopes being $-2.05 \pm 0.32$ ($q=2+\alpha$) and $-2.55 \pm 0.34$ ($q=3+\alpha$). The results in present work suggest that $\log f_{\gamma}^{\rm in}$ and the redshift follow the theoretical relation, which is consistent with that in our previous work (Xiao et al. 2015).

For subclasses of blazars, some similar correlation results are found, but their slopes are different slightly. The slopes of correlations between $\log f_{\gamma}^{\rm in}$ and redshift are $-1.30 \pm 0.47$ (BL Lacs), $-1.34 \pm 0.45$ (FSRQs), $-1.33 \pm 0.44$ (ISP), and $-1.95 \pm 0.44$ (LSP) for the case of $q=2+\alpha$; and those are $-1.65 \pm 0.61$ (BL Lacs), $-1.73 \pm 0.55$ (FSRQs), $-1.67 \pm 0.54$ (ISP), and $-2.51 \pm 0.55$ (LSP) for $q=3+\alpha$. In our results, slopes of BL Lacs are very close to those of FSRQs, and they favor the jet case of $q=3+\alpha$. However, for the whole sample, the results of slopes indicate that two jet cases exist in blazars, since slopes being $-1.82 \pm 0.30$ for $q=2+\alpha$, $- 2.32 \pm 0.37$ for $q=3+\alpha$, and $\sim -2.0$ for theoretical relation.
For the SED classification, results of slopes suggest that ISP blazars favor the jet case of $q=3+\alpha$, while LSP blazars favor the jet case of $q=2+\alpha$. Those results indicate stationary jets ($q=2+\alpha$) are perhaps dominant in LSP blazars.
A possible explanation of those results is that differences of synchrotron peaked frequency caused by the physical differences of blazars, such as the different forms of relativistic jets. In Xiao et al. (2015), we suggested that the continues case ($q=2+\alpha$) of jet is perhaps real for \emph{Fermi} blazars, however, we did not discuss the subclasses at that work.
In the present work, we can not discuss this further since there is no Doppler factor for HSP blazars.

\subsection{Short Variability Time Scale and Luminosity}
For the whole sample of 63 blazars, our results show that 9 blazars violate the E-S or A-N relation in $L_{\gamma}^{\rm ob}$ versus $\Delta T^{\rm ob}$, while the whole sample follows the E-S and A-N relations in $L_{\gamma}^{\rm in}$ versus $\Delta T^{\rm in}$, see Figs. \ref{Lin-2015-intrinsic-logT-logL} and \ref{Lin-2015-intrinsic-logT-logL-BQ}. In our recent work (Xiao et al. 2015), some similar results are found for a sample of 28 blazars.
Different subclasses of blazars have different properties, for instance, FSRQs have strong emission lines, so that can be determined the redshift easier and more accurate than BL Lacs, FSRQs have higher redshift and lower synchrotron peaked frequency than BL Lacs statistically. Therefore, for further analysis, the subclasses of blazars are considered. We find that 9 FSRQs violate the E-S relation and 3 FSRQs violate the A-N relation in $L_{\gamma}^{\rm ob}$ versus $\Delta T^{\rm ob}$, but all blazars follow those relations in their intrinsic properties.
In addition, the averaged values of radio Doppler factors is $< \delta_{\rm R} > \, = \, 6.50 \pm 4.87$ for 32  BL Lacs, and $< \delta_{\rm R} > \, = \, 13.47 \pm 8.04$ for 59 FSRQs. From a K-S test, the probability that the distributions of $\delta_R$ for BL Lacs and FSRQs to be drawn from the same parent distribution is $p = 6.87 \times 10^{-5}$. Thus, the Doppler factors of FSRQs are larger than that of BL Lacs, which is consistent with our previous result (Fan et al. 2004). From above analysis, we find that beaming effect is an important reason that causes the blazars to violate the E-S and A-N relations, and FSRQs have stronger beaming effect than BL Lacs.

\subsection{$\gamma$-ray Emissions and Synchrotron Peaked Frequency}
There is no correlation between $\log f_{\gamma}$ and $\log \nu^{\rm s}_{\rm p}$ with $r=0.07$ and $p=51.51\%$. However, strong positive correlations are found between $\log f^{\rm in}_{\gamma}$ and $\log \nu_{\rm p}^{\rm in}$ for the whole sample. Those correlation coefficients and chance probabilities are $r=0.65$ and $p=2.54 \times 10^{-12}$ for the case of $q=2+\alpha$, and $r=0.68$ and $p=1.97 \times 10^{-13}$ for $q=3+\alpha$. When the redshift effect is removed, strong positive correlations still exist between them. In Lister et al. (2011), strong anti-correlations are found between observed radio flux density at 5GHz and synchrotron peak frequency for BL Lacs and FSRQs. From results in Lister et al. (2011) and this work, the anti-correlations (or no correlation) between observed flux densities and synchrotron peak frequency are significantly different with the positive correlations in intrinsic properties. Thus, the beaming effect can not be ignored when we investigate the physical mechanism of blazars.

 The blazars sequence, which is defined by the anti-correlation between peak luminosity and peak frequency, can be explained by the cooling effect (Fossati et al. 1998; Ghisellini et al. 1998; Wu et al. 2007; Nieppola et al. 2008). However, that theoretical explanation of blazars sequence doesn't consider the beaming effect. Therefore, the intrinsic correlation between peak luminosity and peak frequency is needed to investigate the blazars sequence.
 Wu et al. (2007) estimated Doppler factors ($\delta$) for a sample of 170 BL Lacs, found significant anti-correlations between $\delta$ and $\nu_{\rm p}^{\rm in}$, and between the total 408 MHz luminosity ($L_{\rm 408MHz}$) and $\nu_{\rm p}^{\rm in}$. However, the scatter of $L_{\rm 408MHz}$ versus $\nu_{\rm p}^{\rm in}$ is very large, which is contrast with much tighter relation of blazars sequence. Some similar results are found between radio power and $\nu_{\rm p}^{\rm s}$ in Nieppola et al. (2006). Recently, some high-luminosity high-$\nu_{\rm p}^{\rm s}$ and low-luminosity low-$\nu_{\rm p}^{\rm s}$ sources are detected. Those results indicate that the blazars sequence is likely to be eliminated (Wu et al. 2007).

 Nieppola et al. (2008) collected a sample of 89 AGNs with available Doppler factors, found strong anti-correlation between $\delta$ and $\nu_{\rm p}^{\rm in}$, and proposed that the lower peak frequency blazars are more boosted. In Nieppola et al. (2008), a positive Spearman rank correlation between intrinsic synchrotron peak luminosity ($L_{\rm p}^{\rm in}$) and $ \nu_{\rm p}^{\rm in}$ was also found with $r=0.366$ and $p=3\times10^{-3}$ for blazars, especially for BL Lacs ($r=0.642$ and $p<10^{-3}$). They concluded that the anti-correlation between $L_{\rm p}^{\rm in}$ and $\nu_{\rm p}^{\rm in}$ which is used to determine the blazars sequence is not present, suggesting that the blazar sequence is an artefact of variable Doppler boosting across the peak frequency range. However, the scatter of the correlation between $L_{\rm p}^{\rm in}$ and $ \nu_{\rm p}^{\rm in}$ is about 5 orders of magnitude for their sample.
 In addition, Wu et al. (2009) found a significant positive $L_{\rm p}^{\rm in}$-$\nu_{\rm p}^{\rm in}$ correlation with a Spearman correlation coefficient of $r = 0.59$ at $> 99.99\%$ confidence level. In this work, we find a positive correlation between $\log f^{\rm in}_{\gamma}$ and $\log \nu_{\rm p}^{\rm in}$ after correcting the redshift effect. Thus, our results and previous research indicate that there is a positive correlation between intrinsic emissions and intrinsic synchrotron peak frequency.

 Interestingly, we find a strong positive correlation between $\nu_{\rm p}^{\rm in}$ and $\nu_{\rm p}^{\rm s}$: $\log\nu_{\rm p}^{\rm in} = (1.19 \pm 0.06) \log\nu_{\rm p}^{\rm s} - (3.58 \pm 0.79)$ with $r = 0.91$ and $p = 2.13 \times 10^{-36}$.
 The strong positive $\nu_{\rm p}^{\rm in}$-$\nu_{\rm p}^{\rm s}$ correlation indicates that there is almost no difference of the order of blazars along the peak frequency between before and after considering the beaming effect. Thus the sequence of blazars would not be eliminated in intrinsic properties, although the relation between luminosity and peak frequency is changed significantly. And the observed blazars sequence is highly associated with the intrinsic one.
 Therefore, a new theoretical explanation is needed for the intrinsic blazars sequence.
 In addition, we noticed that the intrinsic blazars sequence could change what we know about blazars, such as the black hole mass difference between BL Lcas and FSRQs.

 The positive correlation between $\gamma$-ray emissions and peak frequency indicates that the synchrotron emissions are highly correlated with $\gamma$-ray emissions. From the synchrotron self-Compton (SSC) process, $\gamma$-ray emissions are produced by the inverse Compton scattering process from synchrotron emissions. So that they should be associated with each other. In addition, we suppose that $f^{\rm in}_{\gamma}$--$\nu_{\rm p}^{\rm in}$ and $\nu_{\rm p}^{\rm in}$--$\nu_{\rm p}^{\rm s}$ relations can be used to estimate the Doppler boosting factors. However, larger sample is needed to find more accurate correlations.

\subsection{Conclusion}
In this work, we collect 91 \emph{Fermi} blazars with available Doppler factors, and investigate the correlations between intrinsic flux density and redshift for the whole sample, BL Lacs, FSRQs, ISP, and LSP respectively. Then, we estimate $\gamma$-ray Doppler factors of 63 blazars, and study the relationship between $\gamma$-ray luminosity and short variability time scale for those blazars. The observed and intrinsic correlations between the $\gamma$-ray flux density and synchrotron peak frequency are also investigated for the whole blazars sample. Our main conclusions are as follows:

1) The correlation between $f^{\rm in}_{\gamma}$ and redshift follows the theoretical relation: $\log f = -2.0 \log z + {\rm const}$.  When the subclasses are considered, we find that the stationary jets are perhaps dominant in LSP blazars.

2) 9 FSRQs violate the Elliot \& Shapiro or the Abramowicz \& Nobili relation in $L ^{\rm ob}_{\gamma}$ versus $\Delta T^{\rm ob}$, while the whole blazar sample obeys to the Elliot \& Shapiro and the Abramowicz \& Nobili relations in $L ^{\rm in}_{\gamma}$ versus $\Delta T^{\rm in}$. Thus FSRQs have stronger beaming effect
than BL Lacs.

3) Strong positive correlation between $f_{\gamma}^{\rm in}$ and $\nu_{\rm p}^{\rm in}$ is found, which suggests that synchrotron emissions are highly correlated with $\gamma$-ray emissions.

\section*{Acknowledgement}

This work is supported by the National Natural Science Foundation of China (U1531245, U1431112, 11203007, 11403006, 10633010, 11173009, 11403006), and the Innovation Foundation of Guangzhou University (IFGZ),
Guangdong Innovation Team (2014KCXTD014),
Guangdong Province Universities and Colleges Pearl River Scholar
Funded Scheme (GDUPS) (2009), Yangcheng Scholar Funded
Scheme (10A027S), and supported for Astrophysics  Key Subjects of Guangdong Province and Guangzhou City.

\clearpage

\clearpage
\begin{center}
\begin{small}
\begin{longtable}{|llclccccc|}
 \caption{The Sample of 91 \emph{Fermi} Blazars.}\\
 \hline
 3FGL name & Other name & z & log$\nu_{\rm p}^{\rm s}$ & Class & $f_{2{\rm GeV}}$ & $\alpha_{\gamma}^{\rm ph}$ & $\delta_{\rm R}$ & Ref \\\hline
 (1) & (2) & (3) & (4) & (5) & (6) & (7) & (8) & (9) \\
 \endfirsthead
 \caption{Continue.}\\
 \hline
 3FGL name & Other name & z & log$\nu_{\rm p}^{\rm s}$ & Class & $f_{2{\rm GeV}}$ & $\alpha_{\gamma}^{\rm ph}$ & $\delta_{\rm R}$ & Ref \\\hline
 (1) & (2) & (3) & (4) & (5) & (6) & (7) & (8) & (9) \\
 \hline
\endhead
 \hline
\endfoot
 \hline
\endlastfoot
 \hline
 J0050.6-0929	&	PKS 0048-09	&	0.300 	&	14.60 	&	IPB	&	12.67 	&	2.09 	&	9.6	&	H09	\\
 J0102.8+5825	&	TXS 0059+581	&	0.643 	&	12.73$^\ast$ 	&	LPQ	&	24.44 	&	2.25 	&	10.91	&	F09	\\
 J0108.7+0134	&	PKS 0106+01	&	2.099 	&	13.53 	&	IPQ	&	28.83 	&	2.39 	&	18.2	&	S10	\\
 J0112.1+2245	&	RX J0112.0+2244	&	0.265 	&	14.39 	&	IPB	&	29.55 	&	2.03 	&	9.1	&	S10	\\
 J0137.0+4752	&	S4 0133+47	&	0.859 	&	12.69 	&	LPQ	&	21.08 	&	2.27 	&	20.5	&	S10	\\
 J0151.6+2205	&	PKS 0149+21	&	1.320 	&	13.14 	&	LPQ	&	2.21 	&	2.65 	&	4.72	&	LV99	\\
 J0205.0+1510	&	4C +15.05	&	0.405 	&	12.10 	&	LPQ	&	2.82 	&	2.53 	&	15.0	&	S10	\\
 J0217.5+7349	&	S5 0212+73	&	2.367 	&	13.35 	&	LPQ	&	7.72 	&	2.91 	&	8.4	&	S10	\\
 J0217.8+0143	&	PKS 0215+015	&	1.721 	&	14.66 	&	IPQ	&	17.03 	&	2.19 	&	5.61	&	F09	\\
 J0222.6+4301	&	3C 66A	&	0.444 	&	14.76 	&	IPB	&	62.03 	&	1.94 	&	2.6	&	H09	\\
 J0237.9+2848	&	B2 0234+28	&	1.207 	&	13.59 	&	LPQ	&	45.20 	&	2.35 	&	16.0	&	S10	\\
 J0238.6+1636	&	PKS 0235+164	&	0.940 	&	13.24 	&	LPB	&	39.92 	&	2.17 	&	23.8	&	S10	\\
 J0303.6+4716	&	4C +47.08	&	0.475 	&	14.10 	&	IPB	&	8.89 	&	2.28 	&	4.33	&	F09	\\
 J0309.0+1029	&	PKS 0306+102	&	0.863 	&	14.04 	&	IPQ	&	11.18 	&	2.23 	&	2.79	&	F09	\\
 J0336.5+3210	&	B2 0333+32	&	1.259 	&	13.55 	&	LPQ	&	5.26 	&	2.89 	&	22.0	&	S10	\\
 J0339.5-0146	&	PKS 0336-01	&	0.852 	&	13.40 	&	LPQ	&	15.32 	&	2.42 	&	17.2	&	S10	\\
 J0423.2-0119	&	PKS 0420-01	&	0.915 	&	12.88 	&	LPQ	&	23.83 	&	2.30 	&	19.7	&	S10	\\
 J0424.7+0035	&	PKS 0422+00	&	1.025 	&	14.22 	&	IPB	&	7.45 	&	2.20 	&	6.11	&	F09	\\
 J0442.6-0017	&	PKS 0440-00	&	0.844 	&	13.04$^\ast$ 	&	LPQ	&	16.65 	&	2.50 	&	12.9	&	H09	\\
 J0449.0+1121	&	PKS 0446+11	&	1.207 	&	13.09 	&	LPQ	&	15.07 	&	2.55 	&	4.90	&	LV99	\\
 J0501.2-0157	&	PKS 0458-02	&	2.286 	&	13.50 	&	IPQ	&	13.33 	&	2.41 	&	15.7	&	S10	\\
 J0522.9-3628	&	PKS 0521-36	&	0.055 	&	13.75 	&	LPQ	&	17.06 	&	2.44 	&	1.83	&	F09	\\
 J0530.8+1330	&	PKS 0528+134	&	2.070 	&	12.53 	&	LPQ	&	22.96 	&	2.51 	&	30.9	&	S10	\\
 J0608.0-0835	&	PKS 0605-08	&	0.872 	&	13.88 	&	IPQ	&	9.27 	&	2.37 	&	7.5	&	S10	\\
 J0721.9+7120	&	1H 0717+714	&	0.310 	&	14.96 	&	IPB	&	75.14 	&	2.04 	&	10.8	&	S10	\\
 J0725.8-0054	&	PKS 0723-008	&	0.127 	&	14.00 	&	IPB	&	3.93 	&	2.19 	&	2.50	&	LV99	\\
 J0738.1+1741	&	PKS 0735+17	&	0.424 	&	14.23 	&	IPB	&	18.41 	&	2.01 	&	3.92	&	F09	\\
 J0739.4+0137	&	PKS 0736+01	&	0.191 	&	14.43 	&	IPQ	&	10.46 	&	2.48 	&	8.5	&	S10	\\
 J0757.0+0956	&	PKS 0754+100	&	0.266 	&	14.05 	&	IPB	&	6.71 	&	2.18 	&	5.5	&	S10	\\
 J0807.9+4946	&	S4 0804+49	&	1.436 	&	13.28 	&	LPQ	&	1.60 	&	2.57 	&	35.2	&	S10	\\
 J0811.3+0146	&	OJ 014	&	0.407 	&	13.28 	&	LPB	&	8.97 	&	2.16 	&	5.39	&	F09	\\
 J0818.2+4223	&	B3 0814+425	&	0.245 	&	13.52 	&	LPB	&	22.22 	&	2.11 	&	4.6	&	S10	\\
 J0820.9-1258	&	PKS 0818-128	&	0.407 	&	14.77 	&	IPB	&	1.03 	&	2.27 	&	3.18	&	F09	\\
 J0830.7+2408	&	B2 0827+24	&	0.941 	&	13.50 	&	LPQ	&	5.79 	&	2.63 	&	13.0	&	S10	\\
 J0831.9+0430	&	PKS 0829+046	&	0.230 	&	13.84 	&	LPB	&	12.43 	&	2.24 	&	3.80	&	F09	\\
 J0841.4+7053	&	RBS 0717	&	2.218 	&	14.44 	&	IPQ	&	11.43 	&	2.84 	&	16.1	&	S10	\\
 J0849.9+5108	&	SBS 0846+513	&	1.860 	&	13.36$^\ast$ 	&	LPB	&	8.79 	&	2.28 	&	6.40	&	LV99	\\
 J0850.2-1214	&	PMN J0850-1213	&	0.566 	&	13.10 	&	LPQ	&	0.27 	&	0.11 	&	16.5	&	H09	\\
 J0854.8+2006	&	PKS 0851+202	&	0.306 	&	14.21 	&	IPB	&	21.46 	&	2.18 	&	16.8	&	S10	\\
 J0948.6+4041	&	B3 0945+408	&	1.249 	&	13.86 	&	IPQ	&	1.81 	&	2.67 	&	6.3	&	S10	\\
 J0956.6+2515	&	OK 290	&	0.712 	&	13.98 	&	IPQ	&	3.95 	&	2.44 	&	4.3	&	H09	\\
 J0957.6+5523	&	4C +55.17	&	0.901 	&	14.74 	&	IPQ	&	34.97 	&	2.00 	&	4.63	&	LV99	\\
 J0958.6+6534	&	S4 0954+65	&	0.367 	&	14.02 	&	IPB	&	5.47 	&	2.38 	&	5.93	&	F09	\\
 J1037.0-2934	&	PKS 1034-293	&	0.312 	&	13.92 	&	IPQ	&	2.03 	&	2.49 	&	2.80	&	F09	\\
 J1058.5+0133	&	PKS 1055+01	&	0.888 	&	13.79 	&	IPQ	&	25.88 	&	2.21 	&	12.1	&	S10	\\
 J1129.9-1446	&	PKS 1127-14	&	1.187 	&	13.99 	&	IPQ	&	5.69 	&	2.79 	&	3.22	&	F09	\\
 J1159.5+2914	&	B2 1156+29	&	0.729 	&	13.04 	&	LPQ	&	32.64 	&	2.21 	&	28.2	&	S10	\\
 J1221.4+2814	&	W Comae	&	0.102 	&	14.83 	&	IPB	&	14.24 	&	2.10 	&	1.2	&	H09	\\
 J1224.9+2122	&	PG 1222+216	&	0.432 	&	14.53 	&	IPQ	&	99.31 	&	2.29 	&	5.2	&	S10	\\
 J1229.1+0202	&	PKS 1226+02	&	0.158 	&	15.12 	&	IPQ	&	36.15 	&	2.66 	&	16.8	&	S10	\\
 J1256.1-0547	&	3C 279	&	0.536 	&	12.69 	&	LPQ	&	83.67 	&	2.34 	&	23.8	&	S10	\\
 J1309.5+1154	&	PKS 1307+121	&	0.407 	&	13.72 	&	LPB	&	2.12 	&	2.14 	&	1.22	&	F09	\\
 J1310.6+3222	&	B2 1308+32	&	0.997 	&	13.22 	&	LPQ	&	15.17 	&	2.25 	&	15.3	&	S10	\\
 J1326.8+2211	&	B2 1324+22	&	1.400 	&	12.97 	&	LPQ	&	8.64 	&	2.45 	&	21.0	&	S10	\\
 J1337.6-1257	&	PKS 1335-12	&	0.539 	&	13.25 	&	LPQ	&	6.63 	&	2.44 	&	8.3	&	S10	\\
 J1408.8-0751	&	PKS B1406-076	&	1.494 	&	12.86 	&	LPQ	&	8.91 	&	2.38 	&	8.26	&	LV99	\\
 J1416.0+1325	&	PKS 1413+135	&	0.247 	&	12.57 	&	LPB	&	2.02 	&	2.36 	&	12.1	&	S10	\\
 J1419.9+5425	&	OQ 530	&	0.151 	&	14.27 	&	IPB	&	3.53 	&	2.31 	&	2.79	&	F09	\\
 J1504.4+1029	&	PKS 1502+106	&	1.839 	&	13.34 	&	LPQ	&	107.57 	&	2.24 	&	11.9	&	S10	\\
 J1512.8-0906	&	PKS 1510-089	&	0.360 	&	13.97 	&	IPQ	&	161.49 	&	2.36 	&	16.5	&	S10	\\
 J1540.8+1449	&	PKS 1538+149	&	0.605 	&	13.97 	&	IPB	&	1.53 	&	2.34 	&	4.3	&	S10	\\
 J1608.6+1029	&	PKS 1606+10	&	1.226 	&	13.39 	&	LPQ	&	7.41 	&	2.62 	&	24.8	&	S10	\\
 J1613.8+3410	&	B2 1611+34	&	1.397 	&	13.44 	&	LPQ	&	3.11 	&	2.35 	&	13.6	&	S10	\\
 J1635.2+3809	&	B3 1633+382	&	1.814 	&	13.21 	&	LPQ	&	60.94 	&	2.40 	&	21.3	&	S10	\\
 J1637.9+5719	&	S4 1637+57	&	0.751 	&	14.22 	&	IPQ	&	1.53 	&	2.81 	&	13.9	&	S10	\\
 J1642.9+3950	&	3C 345	&	0.593 	&	13.46 	&	LPQ	&	10.90 	&	2.45 	&	7.7	&	S10	\\
 J1719.2+1744	&	PKS 1717+177	&	0.407 	&	13.91 	&	IPB	&	5.43 	&	2.04 	&	1.94	&	F09	\\
 J1728.5+0428	&	PKS 1725+044	&	0.293 	&	13.32 	&	LPQ	&	3.65 	&	2.59 	&	3.8	&	H09	\\
 J1733.0-1305	&	PKS 1730-130	&	0.902 	&	12.62 	&	LPQ	&	24.61 	&	2.35 	&	10.6	&	S10	\\
 J1740.3+5211	&	S4 1739+52	&	1.379 	&	13.42 	&	LPQ	&	8.66 	&	2.45 	&	26.3	&	S10	\\
 J1744.3-0353	&	PKS 1741-03	&	1.054 	&	14.06 	&	IPQ	&	2.70 	&	2.27 	&	19.5	&	S10	\\
 J1748.6+7005	&	S4 1749+70	&	0.770 	&	14.27 	&	IPB	&	14.58 	&	2.06 	&	3.75	&	F09	\\
 J1751.5+0939	&	PKS 1749+096	&	0.322 	&	12.99 	&	LPB	&	15.92 	&	2.25 	&	11.9	&	S10	\\
 J1800.5+7827	&	S5 1803+78	&	0.684 	&	13.90 	&	IPB	&	19.78 	&	2.22 	&	12.1	&	S10	\\
 J1806.7+6949	&	3C 371	&	0.051 	&	14.60 	&	IPB	&	11.80 	&	2.23 	&	1.1	&	S10	\\
 J1824.2+5649	&	S4 1823+56	&	0.664 	&	13.25 	&	LPB	&	8.95 	&	2.46 	&	6.3	&	S10	\\
 J1829.6+4844	&	S4 1828+48	&	0.692 	&	13.04$^\ast$ 	&	LPQ	&	6.38 	&	2.37 	&	5.6	&	S10	\\
 J1924.8-2914	&	PKS B1921-293	&	0.352 	&	12.53 	&	LPQ	&	8.84 	&	2.50 	&	9.51	&	F09	\\
 J2005.2+7752	&	S5 2007+77	&	0.342 	&	13.55 	&	LPB	&	6.65 	&	2.22 	&	4.68	&	F09	\\
 J2123.6+0533	&	PKS 2121+053	&	1.941 	&	13.40 	&	LPQ	&	2.02 	&	2.17 	&	15.2	&	S10	\\
 J2134.1-0152	&	PKS 2131-021	&	1.285 	&	13.17 	&	LPB	&	4.62 	&	2.21 	&	7.00	&	F09	\\
 J2147.2+0929	&	PKS 2144+092	&	1.113 	&	13.87 	&	IPQ	&	12.93 	&	2.54 	&	5.96	&	LV99	\\
 J2148.2+0659	&	PKS 2145+06	&	0.990 	&	13.29 	&	LPQ	&	2.30 	&	2.77 	&	15.5	&	S10	\\
 J2158.0-1501	&	PKS 2155-152	&	0.672 	&	13.09 	&	LPQ	& 	4.27 	&	2.27 	&	2.31	&	F09	\\
 J2202.7+4217	&	B3 2200+420	&	0.069 	&	15.10 	&	IPB	&	58.52 	&	2.25 	&	7.2	&	S10	\\
 J2203.7+3143	&	S3 2201+31	&	0.295 	&	14.43 	&	IPQ	&	0.89 	&	3.07 	&	6.6	&	S10	\\
 J2225.8-0454	&	3C 446	&	1.404 	&	13.24 	&	LPQ	&	9.37 	&	2.36 	&	15.9	&	S10	\\
 J2229.7-0833	&	PKS 2227-088	&	1.562 	&	13.34 	&	LPQ	&	20.79 	&	2.55 	&	15.8	&	S10	\\
 J2232.5+1143	&	PKS 2230+11	&	1.037 	&	13.65 	&	LPQ	&	25.54 	&	2.52 	&	15.5	&	S10	\\
 J2236.3+2829	&	B2 2234+28A	&	0.795 	&	12.88 	&	LPB	&	17.70 	&	2.28 	&	6.0	&	H09	\\
 J2254.0+1608	&	PKS 2251+15	&	0.859 	&	13.54 	&	LPQ	&	463.39 	&	2.35 	&	32.9	&	S10	\\
\end{longtable}
\end{small}
\end{center}
{\footnotesize
Note:
 Col. (1) gives the \emph{Fermi} name;
 Col. (2) other name;
 Col. (3) redshift (z);
 Col. (4) synchrotron peak frequency (${\rm log\nu^{s}_{p}}$) in units of Hz from Fan et al. (2016), the data with ``$\ast$" are from 3LAC;
 Col. (5) the classification, which depends on the peak frequency in the sources rest-frame: $\nu_{\rm p}^{\rm res} = (1 + z) \nu_{\rm p}^{\rm obs}$, ``IPQ'' for ISP FSRQs, ``LPQ'' for LSP FSRQ, ``IPB'' for ISP BL Lacs, ``LPB'' for LSP BL Lacs;
 Col. (6) $\gamma$-ray flux density at 2 GeV in units of $10^{-10}$mJy;
 Col. (7) the $\gamma$-ray photon spectral index (${\rm\alpha_{\gamma}^{ph}}$);
 Col. (8) radio Doppler factor ($\delta_{\rm R}$);
 Col. (9) reference of Col. (3) and (8).
Here, F09: Fan et al. (2009); H09: Hovatta et al. (2009); LV99: L\"ahteenim\"aki \& Valtaoja. (1999); S10: Savolainen et al. (2010).
}
\newpage
\clearpage

\begin{center}
\begin{footnotesize}
\begin{longtable}{|llcccccccccccc|}
 \caption{The Short Variability Time Scale and $\gamma$-ray Doppler Factor for \emph{Fermi} Blazars.}\\
 \hline
 3FGL name & Other name & z & Class & log$\Delta T$ & Band & Ref & $F_X$ & Ref & $\alpha_X$ & Ref & $F_{\gamma}$ & $\alpha_{\gamma}^{\rm ph}$ & $\delta_\gamma$ \\\hline
 (1) & (2) & (3) & (4) & (5) & (6) & (7) & (8) & (9) & (10) & (11) & (12) & (13) & (14)\\
 \endfirsthead
 \caption{Continue.}\\
 \hline
 3FGL name & Other name & z & Class & log$\Delta T$ & Band & Ref & $F_X$ & Ref & $\alpha_X$ & Ref & $F_{\gamma}$ & $\alpha_{\gamma}^{\rm ph}$ & $\delta_\gamma$  \\\hline
 (1) & (2) & (3) & (4) & (5) & (6) & (7) & (8) & (9) & (10) & (11) & (12) & (13) & (14)\\
 \hline
\endhead
 \hline
\endfoot
 \hline
\endlastfoot
 \hline
    J0141.4-0929 & 1Jy 0138-097 & 1.034  & B     & 6.03  & $\gamma$ & V13   & 0.70  & LAC   & 1.15  & F14   & 20.65  & 2.12  & 4.45  \\
    J0205.0+1510 & 4C +15.05 & 0.405  & Q     & 5.78  & $\gamma$ & V13   & 0.02  & BZC   & 0.37  & E14   & 6.71  & 2.53  & 0.99  \\
    J0210.7-5101 & PKS 0208-512 & 0.999  & Q     & 5.61  & $\gamma$ & V13   & 1.62  & LAC   & 1.06  & B97   & 47.40  & 2.30  & 5.61  \\
    J0222.6+4301 & 3C 66A & 0.444  & B     & 5.10  & $\gamma$ & V13   & 6.39  & LAC   & 1.60  & F14   & 192.78  & 1.94  & 4.46  \\
    J0238.6+1636 & PKS 0235+164 & 0.940  & B     & 5.94  & $\gamma$ & V13   & 1.24  & BZC   & 1.59  & F14   & 103.05  & 2.17  & 4.41  \\
    J0339.5-0146 & PKS 0336-01 & 0.852  & Q     & 6.02  & $\gamma$ & V13   & 0.75  & BZC   & 0.62  & E14   & 33.58  & 2.42  & 3.67  \\
    J0423.2-0119 & PKS 0420-01 & 0.915  & Q     & 5.37  & $\gamma$ & V13   & 3.87  & LAC   & 0.86  & F14   & 55.92  & 2.30  & 6.84  \\
    J0442.6-0017 & PKS 0440-00 & 0.844  & Q     & 4.95  & $\gamma$ & V13   & 4.06  & LAC   & 0.59  & E14   & 34.94  & 2.50  & 8.05  \\
    J0457.0-2324 & PKS 0454-234 & 1.003  & Q     & 4.83  & $\gamma$ & V13   & 0.60  & BZC   & 0.48  & E14   & 180.35  & 2.21  & 7.31  \\
    J0501.2-0157 & PKS 0458-02 & 2.286  & Q     & 5.73  & $\gamma$ & V13   & 0.92  & BZC   & 0.60  & E14   & 23.24  & 2.41  & 12.30  \\
    J0510.0+1802 & PKS 0507+17 & 0.416  & Q     & 4.03  & $\gamma$ & L15   & 0.38  & BZC   & 0.50  & E14   & 13.34  & 2.41  & 4.34  \\
    J0522.9-3628 & PKS 0521-36 & 0.055  & Q     & 4.57  & $\gamma$ & V13   & 22.50  & LAC   & 0.92  & A09a   & 47.34  & 2.44  & 2.27  \\
    J0530.8+1330 & PKS 0528+134 & 2.070  & Q     & 5.24  & $\gamma$ & D95   & 3.75  & LAC   & 0.58  & F14   & 36.86  & 2.51  & 17.83  \\
    J0538.8-4405 & PKS 0537-441 & 0.894  & B     & 6.04  & $\gamma$ & V13   & 4.53  & LAC   & 1.12  & F14   & 329.60  & 2.04  & 5.36  \\
    J0540.0-2837 & 1Jy 0537-286 & 3.104  & Q     & 6.21  & $\gamma$ & V13   & 1.46  & LAC   & 0.32  & F14   & 7.91  & 2.78  & 16.66  \\
    J0721.9+7120 & 1H 0717+714 & 0.310  & B     & 4.80  & $\gamma$ & V13   & 4.91  & LAC   & 1.77  & F14   & 219.99  & 2.04  & 3.62  \\
    J0738.1+1741 & PKS 0735+17 & 0.424  & B     & 6.05  & $\gamma$ & V13   & 2.09  & LAC   & 1.34  & F14   & 54.82  & 2.01  & 2.69  \\
    J0739.4+0137 & PKS 0736+01 & 0.191  & Q     & 4.86  & Opt   & B83   & 6.36  & LAC   & 0.76  & F14   & 27.34  & 2.48  & 2.94  \\
    J0831.9+0430 & PKS 0829+046 & 0.230  & B     & 6.06  & $\gamma$ & V13   & 0.60  & BZC   & 1.00  & L16   & 33.88  & 2.24  & 1.40  \\
    J0841.4+7053 & RBS 0717 & 2.218  & Q     & 4.41  & $\gamma$ & V13   & 10.70  & LAC   & 0.42  & F14   & 12.58  & 2.84  & 37.76  \\
    J0854.8+2006 & PKS 0851+202 & 0.306  & B     & 5.11  & $\gamma$ & V13   & 1.79  & BZC   & 1.50  & F14   & 59.03  & 2.18  & 2.81  \\
    J0920.9+4442 & S4 0917+44 & 2.189  & Q     & 5.08  & $\gamma$ & V13   & 1.83  & LAC   & 0.39  & F14   & 57.84  & 2.29  & 19.74  \\
    J0957.6+5523 & 4C +55.17 & 0.901  & Q     & 5.50  & $\gamma$ & V13   & 0.77  & LAC   & 0.84  & F14   & 104.50  & 2.00  & 5.10  \\
    J0958.6+6534 & S4 0954+65 & 0.367  & B     & 5.67  & $\gamma$ & V13   & 1.12  & LAC   & 0.24  & F14   & 13.81  & 2.38  & 2.26  \\
    J1104.4+3812 & Mkn 421 & 0.031  & B     & 3.84  & X     & D95   & 678.00  & LAC   & 1.82  & F14   & 302.58  & 1.77  & 4.13  \\
    J1159.5+2914 & B2 1156+29 & 0.729  & Q     & 5.59  & $\gamma$ & V13   & 1.49  & LAC   & 0.86  & F14   & 83.65  & 2.21  & 4.40  \\
    J1217.8+3007 & 1ES 1215+303 & 0.130  & B     & 4.18  & Opt   & G12   & 86.40  & LAC   & 1.47  & B00   & 60.51  & 1.97  & 4.66  \\
    J1221.4+2814 & W Comae & 0.102  & B     & 3.79  & Opt   & F99b  & 2.29  & LAC   & 1.24  & F14   & 41.23  & 2.10  & 2.80  \\
    J1224.9+2122 & PG 1222+216 & 0.432  & Q     & 3.64  & $\gamma$ & V13   & 3.82  & LAC   & 1.19  & F14   & 255.37  & 2.29  & 6.68  \\
    J1229.1+0202 & PKS 1226+02 & 0.158  & Q     & 4.70  & $\gamma$ & V13   & 111.00  & LAC   & 1.11  & F14   & 94.24  & 2.66  & 4.21  \\
    J1256.1-0547 & 3C 279 & 0.536  & Q     & 5.48  & $\gamma$ & V13   & 40.50  & LAC   & 0.84  & F14   & 205.75  & 2.34  & 6.29  \\
    J1310.6+3222 & B2 1308+32 & 0.997  & Q     & 2.65  & Opt   & B83   & 0.85  & LAC   & 0.86  & B97a  & 36.53  & 2.25  & 17.12  \\
    J1408.8-0751 & PKS B1406-076 & 1.494  & Q     & 4.76  & $\gamma$ & F99a  & 0.53  & BZC   & 0.07  & F13   & 17.90  & 2.38  & 12.84  \\
    J1439.2+3931 & PG 1437+398 & 0.344  & B     & 6.25  & $\gamma$ & V13   & 17.90  & LAC   & 1.33  & F14   & 4.44  & 1.77  & 3.25  \\
    J1457.4-3539 & PKS 1454-354 & 1.424  & Q     & 4.64  & $\gamma$ & A09b   & 0.51  & BZC   & 0.68  & E14   & 66.18  & 2.29  & 10.41  \\
    J1504.4+1029 & PKS 1502+106 & 1.839  & Q     & 4.16  & $\gamma$ & A10   & 0.16  & BZC   & 0.84  & F14   & 239.96  & 2.24  & 12.98  \\
    J1512.8-0906 & PKS 1510-089 & 0.360  & Q     & 3.84  & $\gamma$ & V13   & 1.15  & BZC   & 0.98  & F14   & 411.05  & 2.36  & 4.73  \\
    J1517.6-2422 & AP Librae & 0.049  & B     & 2.95  & Opt   & B83   & 2.92  & LAC   & 1.36  & F14   & 52.34  & 2.11  & 2.89  \\
    J1535.0+3721 & RGB J1534+372 & 0.143  & B     & 6.42  & $\gamma$ & V13   & 0.37  & LAC   & 1.84  & F14   & 4.10  & 2.11  & 1.07  \\
    J1540.8+1449 & PKS 1538+149 & 0.605  & B     & 3.44  & Opt   & F96   & 1.82  & LAC   & 0.66  & F14   & 3.70  & 2.34  & 9.88  \\
    J1626.0-2951 & PKS 1622-297 & 0.815  & Q     & 4.14  & $\gamma$ & M97   & 2.28  & LAC   & 0.45  & E14   & 25.02  & 2.45  & 10.66  \\
    J1635.2+3809 & B3 1633+382 & 1.814  & Q     & 4.81  & $\gamma$ & V13   & 0.17  & BZC   & 0.62  & F14   & 114.44  & 2.40  & 10.22  \\
    J1642.9+3950 & 3C 345 & 0.593  & Q     & 5.05  & Opt   & D95   & 4.07  & LAC   & 0.81  & F14   & 25.12  & 2.45  & 5.37  \\
    J1653.9+3945 & Mkn 501 & 0.034  & B     & 5.40  & Inf   & B83   & 65.10  & LAC   & 1.36  & F14   & 97.38  & 1.72  & 1.96  \\
    J1728.3+5013 & I Zw 187 & 0.055  & B     & 5.61  & X     & B83   & 39.60  & LAC   & 1.39  & F14   & 10.81  & 1.96  & 1.86  \\
    J1733.0-1305 & PKS 1730-130 & 0.902  & Q     & 4.90  & $\gamma$ & V13   & 6.32  & LAC   & 0.50  & F14   & 55.89  & 2.35  & 10.02  \\
    J1740.3+5211 & S4 1739+52 & 1.379  & Q     & 5.70  & $\gamma$ & V13   & 1.25  & LAC   & 1.08  & F14   & 16.66  & 2.45  & 6.85  \\
    J1748.6+7005 & S4 1749+70 & 0.770  & B     & 4.68  & $\gamma$ & V13   & 1.55  & LAC   & 1.44  & F14   & 41.38  & 2.06  & 6.14  \\
    J1751.5+0939 & OT 081 & 0.322  & B     & 5.47  & $\gamma$ & V13   & 1.18  & BZC   & 0.74  & L15   & 42.51  & 2.25  & 2.39  \\
    J1800.5+7827 & S5 1803+78 & 0.684  & B     & 4.95  & $\gamma$ & V13   & 1.71  & LAC   & 0.45  & F14   & 50.65  & 2.22  & 5.97  \\
    J1806.7+6949 & 3C 371 & 0.051  & B     & 4.92  & Opt   & B83   & 4.79  & LAC   & 0.75  & F14   & 33.48  & 2.23  & 1.49  \\
    J1813.6+3143 & B2 1811+31 & 0.117  & B     & 6.26  & $\gamma$ & V13   & 1.44  & LAC   & 2.60  & L15   & 15.44  & 2.12  & 1.27  \\
    J1824.2+5649 & S4 1823+56 & 0.664  & B     & 6.60  & $\gamma$ & V13   & 2.52  & LAC   & 0.96  & F14   & 20.11  & 2.46  & 2.86  \\
    J1833.6-2103 & PKS 1830-210 & 2.507  & Q     & 4.44  & $\gamma$ & V13   & 3.25  & LAC   & 0.13  & F14   & 15.44  & 2.12  & 43.49  \\
    J2009.3-4849 & 1Jy 2005-489 & 0.071  & B     & 6.48  & $\gamma$ & V13   & 80.80  & LAC   & 1.32  & F14   & 35.54  & 1.77  & 1.78  \\
    J2134.1-0152 & PKS 2131-021 & 1.285  & B     & 5.88  & $\gamma$ & V13   & 0.67  & LAC   & 1.05  & F14   & 11.18  & 2.21  & 5.66  \\
    J2143.5+1744 & S3 2141+17 & 0.211  & Q     & 5.45  & $\gamma$ & V13   & 1.76  & LAC   & 1.44  & F14   & 44.18  & 2.52  & 1.92  \\
    J2158.8-3013 & PKS 2155-304 & 0.117  & B     & 6.69  & $\gamma$ & V13   & 572.00  & LAC   & 1.62  & F14   & 216.84  & 1.83  & 2.64  \\
    J2202.7+4217 & B3 2200+420 & 0.069  & B     & 4.96  & $\gamma$ & V13   & 7.42  & LAC   & 0.83  & F14   & 164.77  & 2.25  & 1.81  \\
    J2225.8-0454 & 3C 446 & 1.404  & Q     & 3.48  & Opt   & B83   & 2.12  & LAC   & 0.59  & F14   & 19.43  & 2.36  & 22.99  \\
    J2232.5+1143 & PKS 2230+11 & 1.037  & Q     & 4.63  & Opt   & B83   & 3.06  & LAC   & 0.51  & F14   & 50.19  & 2.52  & 11.05  \\
    J2250.1+3825 & B3 2247+381 & 0.119  & B     & 6.15  & $\gamma$ & V13   & 7.93  & LAC   & 1.51  & F14   & 11.02  & 1.91  & 1.69  \\
    J2254.0+1608 & PKS 2251+15 & 0.859  & Q     & 2.94  & $\gamma$ & V13   & 19.00  & LAC   & 0.62  & F14   & 1060.92  & 2.35  & 26.71  \\
\end{longtable}
\end{footnotesize}
\end{center}
{\footnotesize
Note:
Col. (1) gives the \emph{Fermi} name;
Col. (2) other name;
Col. (3) redshift;
Col. (4) classification, ``B" stands for BL Lacs and ``Q" stands for FSRQs;
Col. (5) short variability time scale ($\log \Delta T$) in units of s;
Col. (6) band at which $\Delta T$ is detected;
Col. (7) reference of Col. (3), (5) and (6);
Col. (8) and (9) X-ray flux in units of $10^{-12}$erg/cm$^2$/s at 0.1--2.4 keV and it's reference.
Col. (10) and (11) X-ray spectral index and it's reference;
Col. (12) and (13) $\gamma$-ray integral photon flux at 1--100GeV in units of $10^{-10}$ ph/cm$^2$/s and photon spectrum index ($\alpha_{\gamma}^{\rm ph}$) from 3LAC;
Col. (14) $\gamma$-ray Doppler factor ($\delta_{\gamma}$).
Here, LAC: Ackermann et al. 2015;
BZC: Massaro et al. (2015);
A09a: Ajello et al. (2009);
A09b: Abdo et al. (2009);
A10: Abdo et al. (2010c);
B83: Bassani et al. (1983);
B97: Brinkmann et al. (1997);
B00: Brinkmann et al. (2000);
D95: Dondi \& Ghisellini (1995);
E14: Evans et al. (2014);
F96: Fan et al. (1996);
F99a: Fan et al. (1999a);
F99b: Fan et al. (1999b);
F09: Fan et al. (2009);
F13: Fan et al. (2013b);
F14: Fan et al. (2014a);
G12: Gupta et al. (2012);
H09: Hovatta et al. (2009);
LB15: Liao \& Bai (2015);
LF16: Lin \& Fan (2016);
LV99: L\"ahteenim\"aki \& Valtaoja. (1999);
M97: Mattox et al. (1997);
S10: Savolainen et al. (2010);
V13: Vovk \& Neronov (2013).
If a source has short variability time scales in different references, the value in the latest one is considered.
}

\end{document}